\documentclass[10pt,aps,prl,twocolumn,preprint,nofootinbib]{revtex4}

\usepackage{amsmath}    
\usepackage{graphicx}   
\usepackage{color}
\usepackage{verbatim}

\usepackage{cancel}

\begin{document}

\title{Phenomenological correlations in high-temperature superconductors}
\author{Miguel C N Fiolhais}
\email{miguel.fiolhais@cern.ch}   
\affiliation{Department of Physics, City College of the City University of New York, 160 Convent Avenue, New York 10031, NY, USA}
\affiliation{LIP, Department of Physics, University of Coimbra, 3004-516 Coimbra, Portugal}
\author{Joseph L Birman}
\email{birman@sci.ccny.cuny.edu}   
\affiliation{Department of Physics, City College of the City University of New York, 160 Convent Avenue, New York 10031, NY, USA}

\date{\today}
\begin{abstract}
An interpretation of the quadratic parameter of the Ginzburg-Landau theory of superconductivity is presented in this paper. The negative term in the potential, which allows the spontaneous symmetry breaking, is interpreted as a direct contribution from the energy gap at the Fermi surface to the effective potential. As a result, in the London approximation of the Ginzburg-Landau theory for type-II superconductors, a strong correlation is predicted and observed between the upper critical field at zero kelvin and the critical temperature in high temperature superconductors.
\\
\\
The following paper is published in Europhysics Letters: http://iopscience.iop.org/0295-5075/107/2/27001
\end{abstract}

\maketitle

\section{Introduction}

The Ginzburg-Landau theory of superconductivity at zero kelvin is defined by the following hamiltonian~\cite{ginzburg}:
\begin{eqnarray}
 \mathcal{H} (\psi, {\nabla} \psi, \mathbf{A}, {\nabla} \mathbf{A}) & = &  \frac{1}{2m} \left ( {\nabla} + 2 i e \mathbf{A} \right ) \psi^* \left ({\nabla} - 2 i e \mathbf{A} \right ) \psi \nonumber \\
&+& \alpha |\psi|^2 + \frac{\beta}{2} |\psi|^4  \nonumber \\
&+& \frac{1}{2\mu_0} \left ( {\nabla} \times \mathbf{A} \right )^2 \, ,
\label{hamiltonian}
\end{eqnarray}
with the order parameter $\psi(x) = \rho (x) \textrm{e}^{i\theta (x)}$, where $\rho(x)$ and $\theta (x)$ are real fields, $\mathbf{A}$ is the vector field, $2e$ is the electric charge of the Cooper pairs~\cite{cooper,bardeen,bardeen2}, and the real constants $\alpha$ and $\beta$ are the quadratic and quartic parameters, respectively. In fact, the quartic potential is of the same type that appears in the Higgs mechanism to generate mass terms for gauge bosons in the Standard Model of particle physics~\cite{anderson,englert,higgs,guralnik}, and led to the prediction of a scalar massive particle commonly known as Higgs boson. The recent observation of a Higgs-like signal in the mass range of 124-126 GeV at the Large Hadron Collider is a likely candidate~\cite{atlas,cms}, with no significant deviations from the predicted Standard Model Higgs boson properties observed to date.

In the Higgs mechanism, the potential, $V(\psi) = \alpha |\psi|^2 + \frac{\beta}{2} |\psi|^4 $, must take a negative mass parameter $\alpha$, to generate a non-zero minimum with an infinite number of degenerate states,
\begin{equation}
\langle \psi \rangle^2 = \rho_0^2 = - \frac{\alpha}{\beta} \, ,
\label{eq:london}
\end{equation}
By setting the gauge $\theta (x) = 0$, the symmetry is spontaneously broken, giving rise to a mass term for the vector field, known as the Meissner-Higgs mass~\cite{london,meissner,fiolhais,kleinert3,essen,fiolhaisessen,fiolhais2,fiolhais3},
\begin{eqnarray}
 \mathcal{H} (\psi, {\nabla} \psi, \mathbf{A}, {\nabla} \mathbf{A}) & = &  \frac{1}{2m} \left ( {\nabla} \rho\right )^2 + V (\rho) + \frac{2e^2\rho^2}{m} \mathbf{A}^2 \nonumber \\
&+& \frac{1}{2\mu_0} \left ( {\nabla} \times \mathbf{A} \right )^2 \, .
\label{hamiltonian2}
\end{eqnarray}
This mass term, together with the scalar field mass term, can be associated with two characteristic lengths of a superconductor,
\begin{equation}
\lambda_L = \sqrt{\frac{m}{4\mu_0 e^2 \rho_0^2}} , \,\,\,\,\,\,\,\,\, \xi = \sqrt{\frac{\hbar}{4m|\alpha|}} \, .
\end{equation}
the London penetration length, and the coherence length, respectively.
In fact, the Ginzburg-Landau parameter $\kappa\equiv \lambda_L/\xi$ distinguishes two types of superconductors, $\kappa>1/\sqrt{2}$ for type-I and $\kappa<1/\sqrt{2}$, for type-II superconductors~\cite{halperin,kleinert,kleinert2,fiolhaiskleinert}. 
Moreover, the condensation energy, \emph{i.e.} the necessary energy to restore the vacuum symmetry, also allows interesting properties to unfold, such as the thermodynamic critical magnetic field,
\begin{equation}
B_c = \frac{1}{4}\frac{\hbar}{e}\frac{1}{\lambda_L\xi} \, .
\end{equation}
In the case of type-II superconductors, the thermodynamic critical field relates to the upper field as,
\begin{equation}
B_{c2} = \sqrt{2} \frac{\lambda_L}{\xi} B_{c} = \frac{\sqrt{2}}{4}\frac{\hbar}{e}\frac{1}{\xi^2} \, .
\end{equation}

\begin{figure}[!ht]
\begin{center}
\includegraphics[height=18.cm]{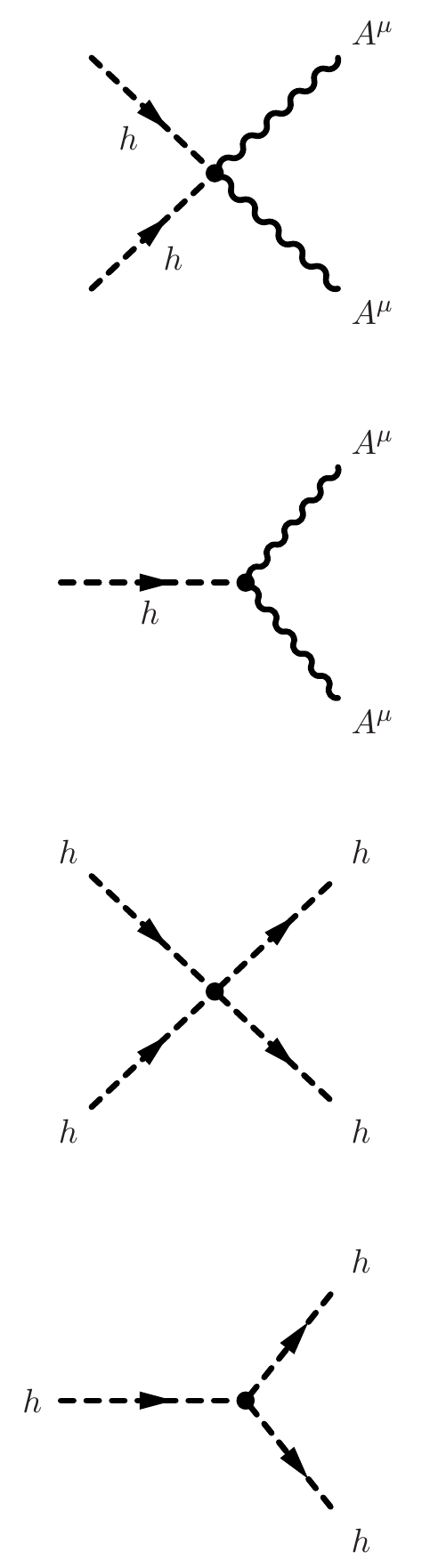}
\caption{Interaction vertices of the Higgs field and the massive gauge field $A^\mu$ at tree level.}
\label{feynmandiagrams}
\end{center}
\end{figure}

As in the electroweak interactions, the scalar field can be expanded as,
\begin{equation}
\rho (x) = \rho_0 + h(x) \, ,
\end{equation}
where $h(x)$ is the Higgs field corresponding to the fluctuations around the ground state. In the context of a (3+1)-dimensional scalar electrodynamical Higgs model, possessing canonical commutation rules, this scalar field can be regarded as a particle~\cite{altland}, as the Higgs boson in the electroweak theory. This concept led to the discovery of pions as the quantum of the forces of nuclear interactions in the first half of the twentieth century, and instigated the search for a Higgs boson during several decades in a number of particle physics accelerators. As such, the hamiltonian (\ref{hamiltonian2}) written in terms of the Higgs field,
\begin{eqnarray}
 \mathcal{H} (h, {\nabla} h, \mathbf{A}, {\nabla} \mathbf{A}) 
& = &  \frac{1}{2m} \left ( {\nabla} h \right )^2 +  \frac{2e^2\rho_0^2}{m} \mathbf{A}^2 \nonumber \\ 
& + & \frac{2e^2}{m} h^2 \mathbf{A}^2 + \frac{4e^2\rho_0}{m} h \mathbf{A}^2 \nonumber \\ 
& + & 2 \beta \rho_0 ^2 h^2 + 2 \beta \rho_0 h^3 + \frac{1}{2}\beta h^4 \nonumber \\
& + & \frac{1}{2\mu_0} \left ( {\nabla} \times \mathbf{A} \right )^2 \, ,
\end{eqnarray}
gives rise to several interaction vertices at tree level, depicted in Figure~\ref{feynmandiagrams}. The meaning of such Higgs field in the case of superconductivity and, in particular, the physical interpretation of its mass, will be discussed ahead.

The Ginzburg-Landau theory can be nowadays regarded as a predecessor of what is now called (3+1)-dimensional scalar quantum electrodynamics, that was
studied in detail by Coleman and Weinberg~\cite{coleman,weinberg}. 
The multiple successes of the theory, from the description of the phase transition to the prediction of a coherence length or quantum vortices~\cite{abrikosov}, have established it as the main macroscopic quantum theory of superconductivity. However, there are still a few details open to discussion, such as the physical interpretation of its phenomenological parameters, which can unfold interesting properties.
It should be stressed, however, that the Ginzburg-Landau theory, in the London approximation, shall only be applied to type-II superconductors with large GL parameters, also known as ``clean'' superconductors. In type-I superconductors, the penetration length is of the same order of magnitude or smaller than the coherence length, yielding non-local effects, and therefore, the Pippard's model must be taken into account~\cite{pippard}.


\section{Correlation between Upper Critical Field and Critical Temperature}

In the Ginzburg-Landau theory, the pairing energy, or energy gap, resulting from the attractive potential binding electrons together, is not explicitly included, as the phenomenological constants are free, and fixed by the experiment. In fact, both the quadratic and quartic parameters can be microscopically derived from BCS theory near the phase transition in terms of the critical temperature and the Fermi energy~\cite{gorkov},
\begin{equation}
\alpha = - \frac{6\pi^2(kT_c)^2}{7\zeta(3)\epsilon_F^0} \left ( 1- \frac{T}{T_c} \right )\, , \,\,\,\,\, \beta = - \frac{6\pi^2(kT_c)^2}{7\zeta(3)\epsilon_F^0\rho_0^2} \, ,
\end{equation}
where $\zeta$ is the zeta function, and $\epsilon_F^0$ is the Fermi energy.
However, assuming the BCS relation for the coherence length, the Fermi velocity, and the energy gap, this result is consistent with the Ginzburg-Landau theory predictions, and therefore, no additional information can be extracted from it. {The quadratic term vanishes as the superconducting gap converges to zero
at the critical temperature, and so does the density of pairs. In other words, the vacuum expectation value of the quartic 
potential becomes zero and the symmetry is restored.}

In this paper, the negative contribution to the effective potential is assumed to be directly related to the energy gap, leading to important implications. {The main motivation for this assumption lies in the fact that the pairing energy should be the only source of a negative energy density in the macroscopic Ginzburg-Landau theory, proportional to the density of pairs.} As such, the effective potential becomes,
\begin{equation}
V(\rho) = -2 \Delta \rho^2 + \frac{1}{2}\beta\rho^4 \, ,
\end{equation}
where the first term corresponds to the energy density resulting from the binding energy. It should be stressed, however, that this assumption is only valid at zero kelvin. At a finite temperature new quadratic and quartic contributions arise, and such conjecture is no longer reasonable.

\begin{figure}[!ht]
\begin{center}
\includegraphics[height=6.0cm]{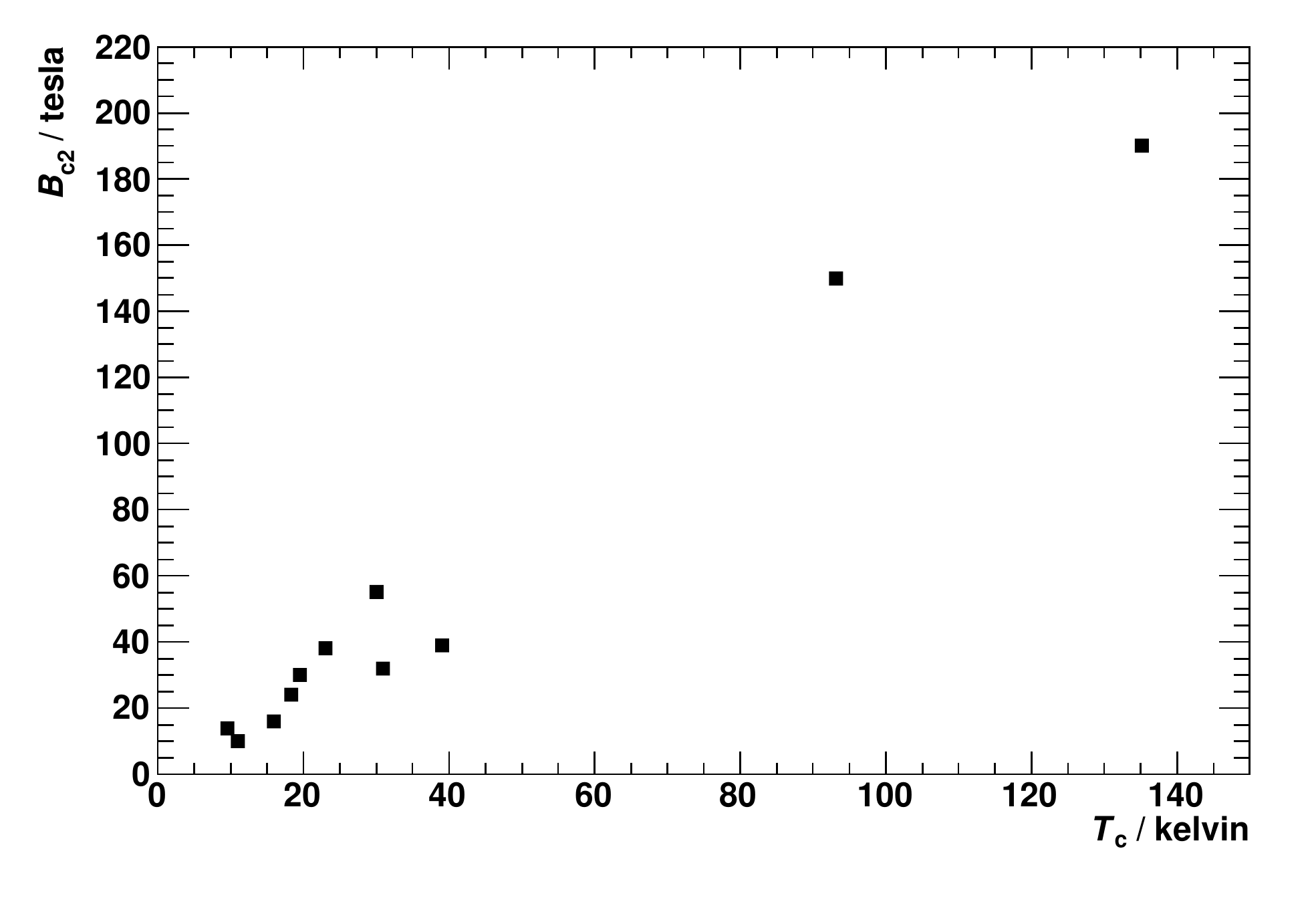}
\caption{Correlation between the upper critical field and the critical temperature for several clean high-temperature superconductors.}
\label{graphic}
\end{center}
\end{figure}

As in high-temperature superconductors the pairing usually holds a \emph{d}-wave symmetry, the mean-field BCS theory for \emph{d}-wave superconductors predicts an energy gap proportional to the critical temperature,
\begin{equation}
\Delta = 2.14 k T_c \, .
\end{equation}
Despite the observed violation of the BCS result, especially in underdoped cuprates, data from ARPES and tunnel spectroscopy reveal that the superconducting energy gap magnitude at the gap nodes is in fact proportional to the critical temperature for both overdoped and underdoped high temperature superconductors, in agreement with the mean-field BCS scaling law. A detailed discussion on this topic by Panagopoulos and Xiang can be found in~\cite{panagopoulos}.
In this scenario, a relation between the upper critical field and the critical temperature is, therefore, straightforward,
\begin{equation}
B_{c2} = \frac{2.14}{\sqrt{2}}\frac{k m}{\hbar e} T_c \, ,
\end{equation}
where 
\begin{equation}
\frac{2.14}{\sqrt{2}}\frac{k m}{\hbar e} \approx 1.13~\textrm{T/K} \, .
\end{equation}
The predicted linearity between the upper critical field and the critical temperature for clean high-temperature superconductors can be observed in Figure~\ref{graphic} for several materials. The experimental data points were extracted from~\cite{mourachkine}, and only superconductors with $k \ge 100$ were considered. The strong correlation observed corroborates the predicted relation between the two observables. However, it cannot be regarded as a fundamental relation of superconductivity, but more as a phenomenological correlation, due to the difficulties mentioned aboved and the variety of symmetries in superconductors pairing. Furthermore, no quantum corrections to the quartic potential were taken into account in these calculations, as only classical tree level contributions were considered.

In this model, the scalar field, $h(x)$, permeates the superconducting bulk, giving mass to photons, and therefore, suppressing the electromagnetic interaction inside. This massive scalar field corresponds to a collective excitation of the Cooper pairs in the lattice, which have a kinetic mass of  $2m$. Such kinetic mass competes directly with the energy gap resulting from the attractive potential between the pairing electrons. However, since the electron mass is much larger than the pairing energy, its contribution is usually neglected. On the other hand, the negative component of the effective potential, interpreted here as a contribution from the binding energy, can be directly related to the mass of the scalar Higgs-like field,
\begin{equation}
m_h = \frac{1}{2} \frac{\partial^2 V}{\partial h^2} \bigg|_{h = 0} = \rho_0^2 \beta = |\alpha| = 2\Delta \, .
\end{equation}

\section{Summary}

In this letter, the negative quadratic contribution to the effective potential of the Ginzburg-Landau theory, at zero kelvin, is conjectured to be directly due to the pairing energy between electrons. Such assumption leads to correlations between observables for clean high-temperature superconductors, corroborated with experimental data. Furthermore, the meaning of the Higgs field and its mass inside the superconductor are also discussed in the context of (3+1)-dimensional scalar quantum electrodynamics, allowing for a quasi-particle interpretation.



\end{document}